\documentclass[aip,sd,amsmath,amssymb, reprint,]{revtex4-1}
\usepackage{graphicx}
\usepackage{dcolumn}
\usepackage{bm}

\begin{document}

\title{Quantum delocalization of protons in the hydrogen bond network of an enzyme active site}

\author{Lu Wang, Stephen D. Fried, Steven G. Boxer and Thomas E. Markland}
\affiliation{Department of Chemistry, Stanford University, 333 Campus Drive, Stanford, California 94305}
\email{tmarkland@stanford.edu}

\date{\today}

\begin{abstract}
Enzymes utilize protein architectures to create highly specialized structural motifs that can greatly enhance the rates of complex chemical transformations. Here we use experiments, combined with {\it{ab initio}} simulations that exactly include nuclear quantum effects, to show that a triad of strongly hydrogen bonded tyrosine residues within the active site of the enzyme ketosteroid isomerase (KSI) facilitates quantum proton delocalization. This delocalization dramatically stabilizes the deprotonation of an active site tyrosine residue, resulting in a very large isotope effect on its acidity. When an intermediate analog is docked, it is incorporated into the hydrogen bond network, giving rise to extended quantum proton delocalization in the active site. These results shed light on the role of nuclear quantum effects in the hydrogen bond network that stabilizes the reactive intermediate of KSI, and the behavior of protons in biological systems containing strong hydrogen bonds.
\end{abstract}

\maketitle

\section{\label{sec:level1}Introduction}
While many biological processes can be well described with classical mechanics, there has been much interest and debate as to the role of quantum effects in biological systems ranging from photosynthetic energy transfer, to photoinduced isomerization in the vision cycle and avian magnetoreception \cite{lambert2013}. For example, nuclear quantum effects, such as tunneling and zero-point energy (ZPE), have been observed to lead to {\it kinetic} isotope effects of greater than 100 in biological proton and proton-coupled electron transfer processes \cite{sutcliffe2002,klinman2013}. However, the role of nuclear quantum effects in determining the ground state thermodynamic properties of biological systems, manifesting as {\it equilibrium} isotope effects, has gained significantly less attention \cite{perez2010}.

\begin{figure}
\centering
\includegraphics[width=0.8\columnwidth]{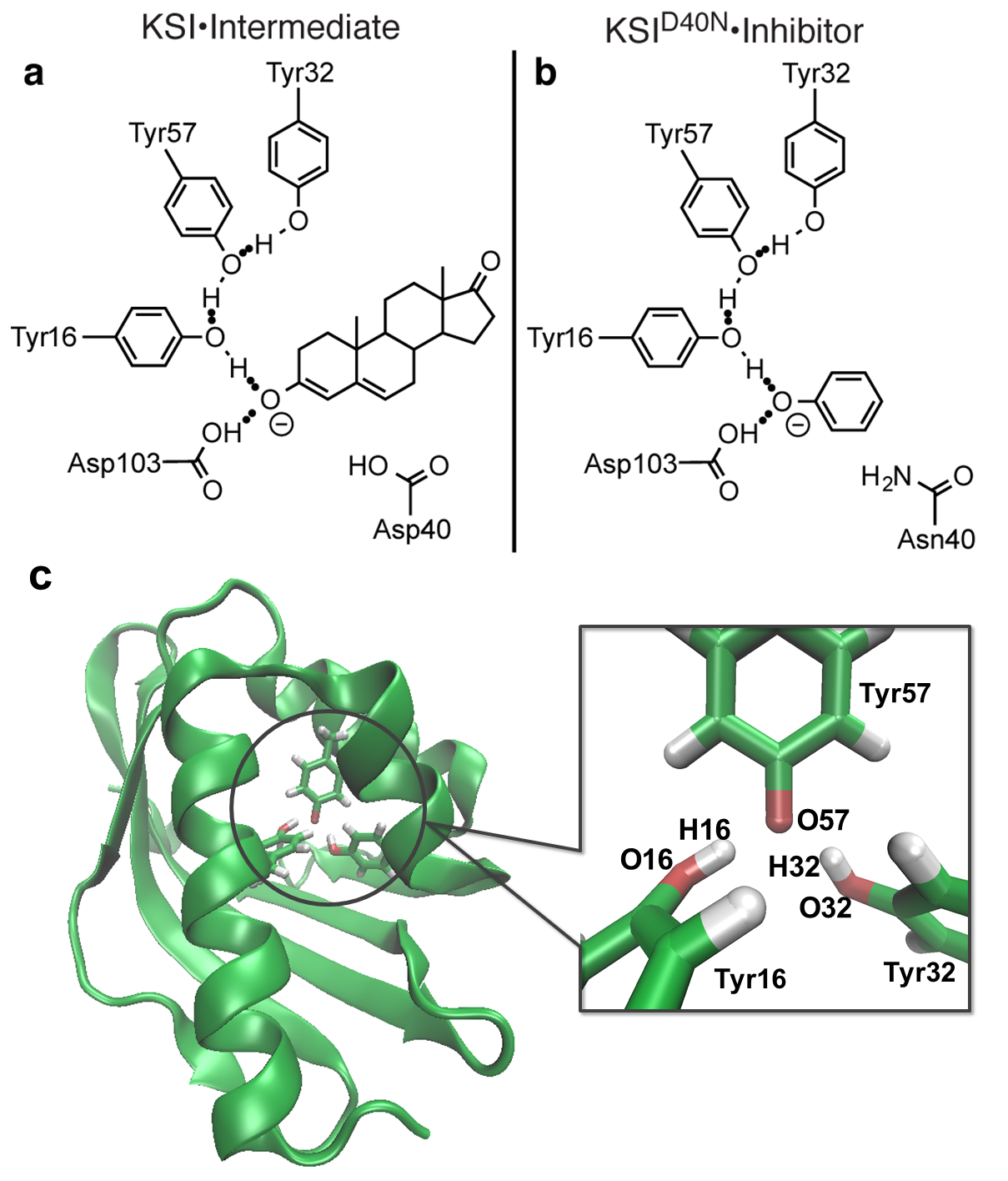}
\caption{ KSI$\cdot$intermediate and KSI$^{D40N}\cdot$inhibitor complex. Schematic depiction of (a) the KSI$\cdot$intermediate complex during the catalytic cycle (Fig. S1) and (b) a complex between KSI$^{D40N}$ and phenol, an inhibitor which acts as an intermediate analog. Both the intermediate and inhibitor are stabilized by a hydrogen bond network in the active site of KSI. (c) Image of KSI$^{D40N}$ with the tyrosine triad enlarged and the atoms O16, H16, O32, H32 and O57 labeled (shown with Tyr57 deprotonated) \cite{Ha2000}.}
\label{fig:fig1}
\end{figure}

Ketosteroid isomerase (KSI) possesses one of the highest enzyme unimolecular rate constants and is thus considered a paradigm of proton transfer catalysis in enzymology \cite{feierberg2002,Pollack2004,Warshel2007,Chakravorty2010,Hanoian2010,Herschlag2013,amyes2013}. KSI's remarkable rate is intimately connected to the formation of a hydrogen bond network in its active site (Fig. 1a) which acts to stabilize a charged dienolate intermediate, lowering its free energy by $\sim$11 kcal/mol relative to solution (Fig. S1) \cite{Pollack2004}. This extended hydrogen bond network in the active site links the substrate to Asp103 and Tyr16, with the latter further hydrogen bonded to Tyr57 and Tyr32 as shown in Fig. 1a.

The mutant KSI$^{D40N}$ preserves the structure of the wild-type enzyme while mimicking the protonation state of residue 40 in the intermediate complex (Fig. 1b), therefore permitting experimental investigation of an intermediate-like state of the enzyme \cite{Pollack2004,Fafarman2012,Fried2013,Sigala2013}. Experiments have identified that, in the absence of an inhibitor, one of the residues in the active site of KSI$^{D40N}$ is deprotonated \cite{Fafarman2012}. Although one might expect the carboxylic acid of Asp103 to be deprotonated, the combination of recent $^{13}$C NMR and UV-Vis experiments has shown that the ionization resides primarily on the hydroxyl group of Tyr57, which possesses an anomalously low $pK_a$ of 6.3 $\pm$ 0.1 \cite{Fafarman2012}. Such a large tyrosine acidity is often associated with specific stabilizing electrostatic interactions (such as a metal ion or cationic residue in close proximity), which is not the case here suggesting an additional stabilization mechanism is at play \cite{schwans2013}.

One possible explanation is suggested by the close proximity of the oxygen atoms (O) on the side chains of the adjacent residues Tyr16 (O16) and Tyr32 (O32) to the deprotonated O on Tyr57 (O57, Fig. 1c) \cite{Ha2000}. In several high-resolution crystal structures, these distances are found to be around 2.6~\AA~  \cite{Ha2000,Sigala2008,Sigala2013}, which are much shorter than those observed in hydrogen bonded liquids such as water, where O--O distances are typically around 2.85 \AA. Such short heavy atom distances are only slightly larger than those typically associated with low-barrier hydrogen bonds (LBHB) \cite{Cleland1994,Frey1994,Zhao1996}, where extensive proton sharing is expected to occur between the atoms. In addition, at these short distances the proton's position uncertainty (de Broglie wavelength) becomes comparable to the O--O distance, indicating that nuclear quantum effects could play an important role in stabilizing the deprotonated residue (Fig 1c). In this work, we demonstrate how nuclear quantum effects determine the properties of protons in KSI$^{D40N}$'s active site hydrogen bond network in the absence and presence of an intermediate analog by combining  {\small{\it{ab initio}}} path integral simulations and isotope effect experiments.

\section{\label{sec:level1}Results and discussions}
\subsection{\label{sec:level2}Isotope substitution experiments reveal large isotope effect on acidity}
To assess the impact of nuclear quantum effects on the anomalous acidity of Tyr57, we measured the isotope effect on the acid dissociation constant upon substituting hydrogens (H) in the hydrogen bond network with deuterium (D). Because tyrosinate absorbs light at 300 nm more intensely than tyrosine, titration curves were generated by recording UV spectra of KSI$^{D40N}$ at different $pL$'s (where {\it{L}} is H or D) \cite{schwans2013}. These experiments (Fig. 2 and Table S1) reveal a change in $pK_a$ upon H/D substitution (${\Delta}pK_a^{KSI}$) of 1.1 $\pm$ 0.14 for tyrosine in the KSI$^{D40N}$ active site. This $pK_a$ isotope effect is much larger than that observed for tyrosine in solution (${\Delta}pK_a^{Sol}$ = 0.53 $\pm$ 0.08), and is also, to the best of our knowledge, the largest recorded $pK_a$ isotope effect \cite{wolfsberg2009}. Changes in static equilibrium properties, such as the $pK_a$, upon isotope substitution arise entirely from the quantum mechanical nature of nuclei. Such a large excess isotope effect, defined as $\Delta{\Delta}pK_a\equiv{\Delta}pK_a^{KSI}-{\Delta}pK_a^{Sol}$, of 0.57 $\pm$ 0.16 thus indicates that the tyrosine triad in the active site of KSI$^{D40N}$ exhibits much larger nuclear quantum effects than those observed for tyrosine in aqueous solution.

\begin{figure}
\centering
\includegraphics[width=0.9 \columnwidth]{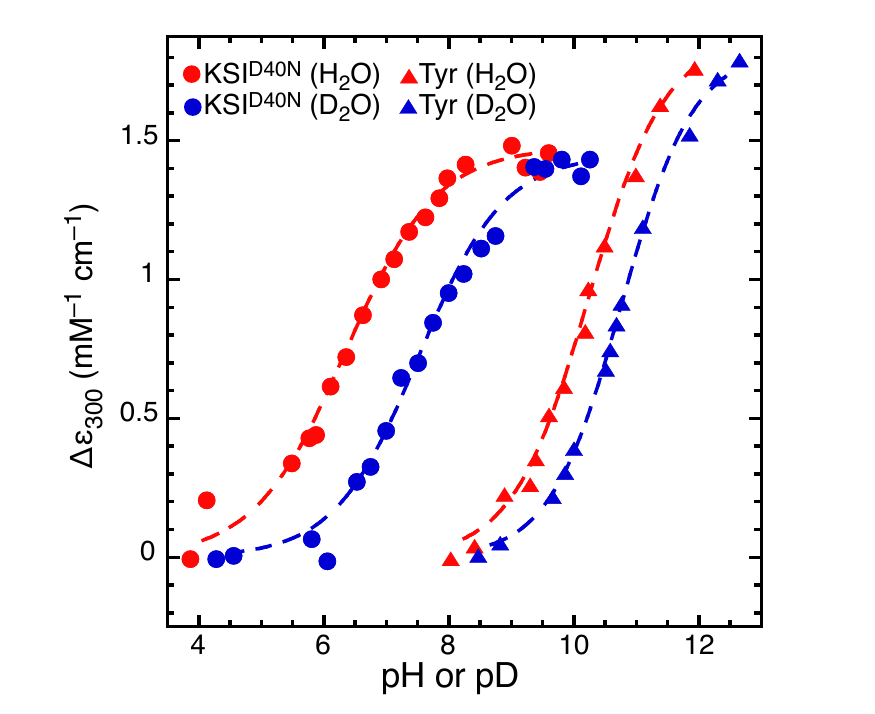}
\caption{Experimental UV-Vis titration curves of KSI$^{D40N}$ (circles) and tyrosine (triangles) in buffered H$_2$O (red) and D$_2$O (blue). Fractional ionization of the phenolic moiety is monitored by measuring the change in absorption at 300 nm. All data sets were well fit to a sigmoid function, admitting four $pK_a$ values and two $\Delta pK_a$ values, the acidity isotope effect.}
\label{fig:fig2}
\end{figure}

Given the possible role of nuclear quantum effects in KSI$^{D40N}$, can one estimate how much the quantum nature of protons changes the acidity of Tyr57 compared to a situation in which all the nuclei in the enzyme active site were classical ($m\rightarrow\infty$)? In the quasi-harmonic limit one can show that the $pK_a$ varies as the inverse square root of the particle mass (m) \cite{ceri-mark13jcp,marsalek2014}. By using this relation and the experimental $pL$ values, we can extrapolate to the classical ($m\rightarrow\infty$) limit which yields that the $pK_a$ of Tyr57 in KSI$^{D40N}$ would be 10.1 $\pm$ 0.5 if the hydrogens were classical particles (Fig. S2). Relative to the observed $pK_a$ of 6.3 $\pm$ 0.1, this implies nuclear quantum effects lower the $pK_a$ or Tyr57 by 3.8 $\pm$ 0.5 units: an almost four orders of magnitude change in the acid dissociation constant.

\subsection{\label{sec:level2}{\textit{Ab initio}} path integral simulations of KSI$^{D40N}$}
To provide insights into the molecular origins of the nuclear and electronic quantum effects which stabilize the deprotonated Tyr57 residue, we performed simulations of KSI$^{D40N}$. To treat the electronic structure in the active site we performed {\it{ab initio}} molecular dynamics (AIMD) simulations using a QM/MM approach \cite{warshel1976,Field1990,Eurenius1996,monard1999} in which the QM region was treated by density functional theory at the B3LYP-D3 level \cite{beck93jcp,grim+10jcp} (see Methods). These simulations allow for bond breaking and formation as dictated by the instantaneous electronic structure rather than pre-defined bonding rules.

AIMD simulations are typically performed treating the nuclei as classical particles. However, a classical treatment of the nuclei would predict that the $pK_a$ would not change upon isotope substitution. Nuclear quantum effects can be exactly included in the static equilibrium properties for a given description of the electronic structure using the path integral formalism of quantum mechanics, which exploits the exact mapping of a system of quantum mechanical particles onto a classical system of ring polymers \cite{feyn-hibb65book,chan-woly81jcp,bern-thir86arpc,marx-parr95jcp}. We combined this formalism with on-the-fly electronic structure calculations and performed {\it{ab initio}} path integral molecular dynamics (AI-PIMD) simulations of KSI$^{D40N}$. These simulations treat both the nuclear and electronic degrees of freedom quantum mechanically in the active site QM region, and also incorporate the fluctuations of the protein and solvent environment in the MM region. The simulations consisted of between 47 and 68 QM atoms and more than 52,000 MM atoms describing the rest of the protein and solvent (Table S2). These simulations, which until recently would have been computationally prohibitive, were made possible by accelerating the PIMD convergence using a generalized Langevin equation \cite{ceri-mano12prl}, utilizing new methods to accelerate the extraction of isotope effects \cite{ceri-mark13jcp} and exploiting graphical processing units (GPUs) to perform efficient electronic structure theory evaluations via an interface to the TeraChem code \cite{ufimtsev2009,isborn2012}. Such a combination yielded almost three orders of magnitude speed-up compared to existing AI-PIMD approaches, allowing 1.1 ps/day of simulation to be obtained using 6 NVIDIA Titan GPUs. We have recently shown that AI-PIMD simulations using the B3LYP-D3 functional give excellent predictions of isotope effects in water, validating such a combination for the simulation of isotope effects in hydrogen bonded systems \cite{wang2014}.

\begin{figure*}
\centering
\includegraphics[height=8.5cm]{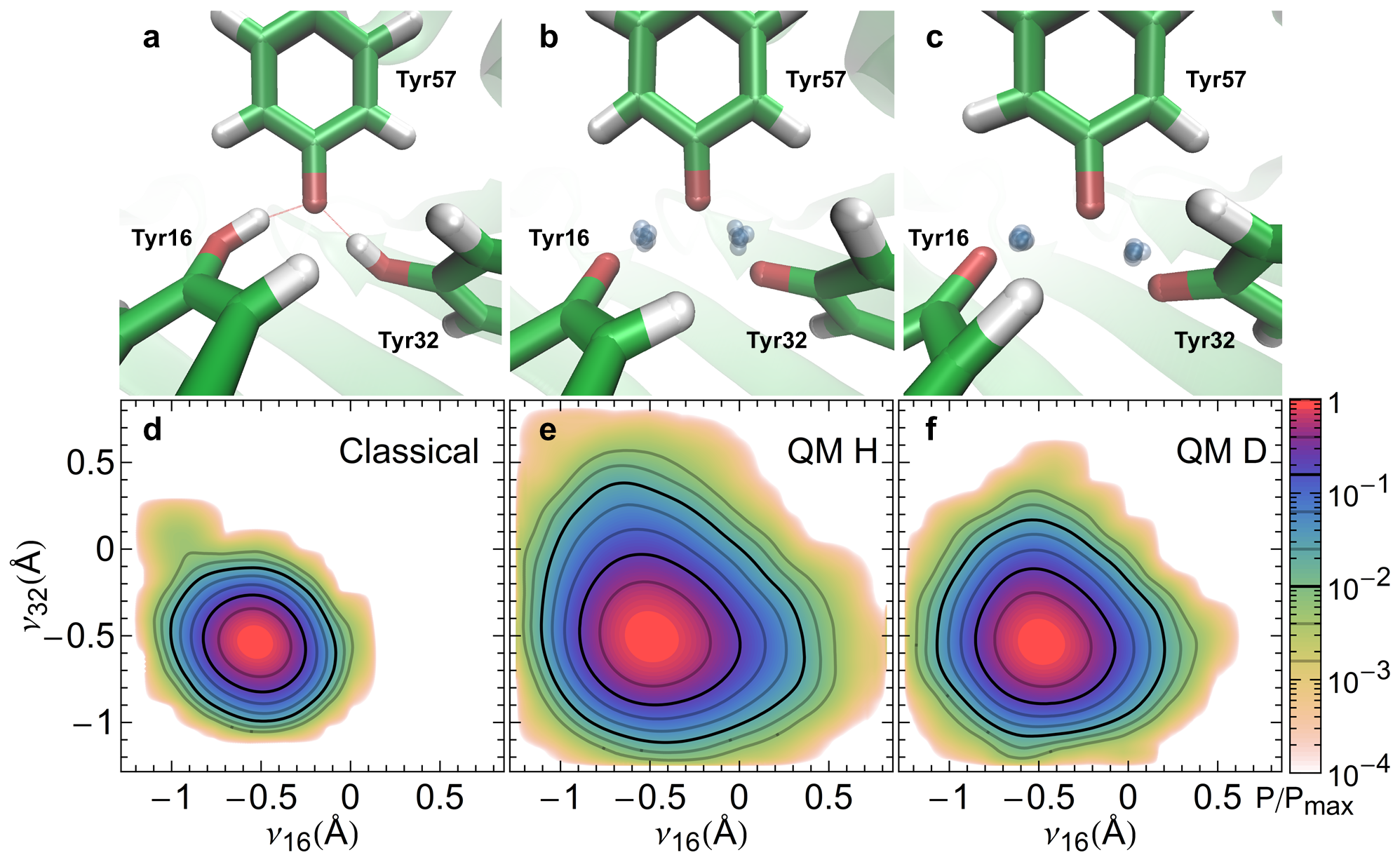}
\caption{Delocalized protons in the active site of KSI$^{D40N}$ from AIMD and AI-PIMD simulations. Snapshots of the active site of KSI$^{D40N}$ (top panels) and probability distribution along the proton sharing coordinates $\nu_{16}$ and $\nu_{32}$ (bottom panels) when the nuclei are treated classically (Classical) or quantum mechanically for H (QM H) and D (QM D). In the top panels, green, red and white represent C, O and H atoms. The blue-grey spheres in the QM snapshots show uncertainty in the delocalized protons’ positions. For clarity all other particles are represented by their centroids. In the bottom panels, probabilities are shown on a log scale and are normalized by their maximum values.}
\label{fig:fig3}
\end{figure*}

\subsection{\label{sec:level2}Quantum delocalization of protons in KSI$^{D40N}$}
The excess isotope effect, $\Delta\Delta pK_a$, obtained from our simulations (see SI Methods Ic) was 0.50 $\pm$ 0.03, in excellent agreement with the experimental value of 0.57 $\pm$ 0.16. The average distance between O57 and the adjacent O16 and O32 atoms obtained in our simulations were 2.56 \AA ~and 2.57 \AA, with a standard deviation in both cases of 0.09 \AA. The distribution of distances between O16 and O57 explored in the simulation is shown in the Fig. 4a inset. These average O--O distances are slightly smaller than (and within the margin of error of) those in the starting crystal structure ($\sim$ 2.6 \AA) \cite{Ha2000}. As we will discuss below, the close proximity of the neighboring O16 and O32 groups plays a crucial role in the origins of the observed isotope effect.

Figure 3a-c shows snapshots from AIMD simulations in which the nuclei are treated classically (Fig. 3a) or quantum mechanically using the path integral formalism (AI-PIMD, Fig. 3b and c), while videos of the simulation trajectories are provided in SI Videos 1-3. For the quantum simulations the H16 and H32 protons are shown as their full ring polymers, which arise from the path integral quantum mechanics formalism. The spread of the ring polymer representing each proton is related to its de Broglie wavelength (quantum mechanical position uncertainty) \cite{parrinello1984,mark-bern12pnas}. The uncertainty principle dictates that localization of a quantum mechanical particle increases its quantum kinetic energy. The protons will thus attempt to delocalize, i.e. spread their ring polymers, to reduce this energetic penalty. The resulting proton positions in Fig. 3b arise from the interplay between the chemical environment, such as the covalent O--H bond, which acts to localize the proton and the quantum kinetic energy penalty that must be paid to confine a quantum particle. Inclusion of nuclear quantum effects thus allows the protons to delocalize between the hydroxyl oxygens to mitigate the quantum kinetic energy penalty (Fig. 3b), which is not observed classically (Fig. 3a). Confinement of D, which due to its larger mass has a smaller position uncertainty, leads to a much less severe quantum kinetic energy penalty and hence less delocalization (Fig. 3c).

To characterize the degree of proton delocalization we define a proton sharing coordinate $\nu_X=d_{OX,HX}-d_{O57,HX}$, where $d_{OX,HX}$ is the distance of proton HX from oxygen atom OX and X=16 or 32. Hence $\nu_X=0$ corresponds to a proton that is equidistant between the oxygen atoms of TyrX and Tyr57, while a positive value indicates proton transfer to Tyr57 from TyrX. Figures 3 d-f show the probability distribution along the proton sharing coordinates $\nu_{16}$ and $\nu_{32}$ for classical nuclei and quantum nuclei for H and D, respectively. The free energies along $\nu_{16}$ and $\nu_{32}$ are provided in Fig. S3. In the classical AIMD simulation, H16 and H32 remain bound to their respective oxygens throughout the simulation ($\nu_{16}$ and $\nu_{32}$ are negative) with Tyr57 ionized 99.96\% of the time (Fig. 3d). However, upon including nuclear quantum effects (AI-PIMD simulations) there is a dramatic increase in the range of values $\nu_{16}$ and $\nu_{32}$ can explore (Fig. 3e). In particular, the probability that Tyr57 is protonated ($\nu_X > 0$) increases by about 150-fold for H upon including quantum effects (Fig. 3e), with the proton ``hole'' equally shifted onto the adjacent Tyr16 or Tyr32 residues. Proton transfers between the residues are observed frequently (SI Videos 2 and 3) with site lifetimes on the order of 60 and 200 fs in the H and D simulations, respectively. Although PIMD simulations exactly include nuclear quantum effects for calculating static properties, they do not allow rigorous extraction of time-dependent properties; nevertheless, they offer a crude way to assess the time-scale of the proton motion. The frequent transfers observed are also consistent with Fig. 3e and f, which show a monotonic decrease in the probability along both $\nu_{16}$ and $\nu_{32}$, i.e. although the proton transferred state is lower in probability, the proton transfer process along each of the proton sharing coordinates contains no free energy barrier (Fig. S3) and is thus kinetically fast. 

As an experimental counterpart, we used chemical shifts of $^{13}$C$_{\zeta}$-Tyr labeled KSI$^{D40N}$ as a measure of fractional ionization of each Tyr residue (SI Methods Ib) \cite{Sigala2013}. This analysis yielded values of 79\% for the Tyr57 ionization for H and 86\% for D compared with simulated values of 94.2\% and 98.3\% ($\pm$ 0.3\%), respectively. This represents good quantitative agreement, since the population difference amounts to a difference in the relative free energy between experiment and theory of 0.7 kcal/mol -- an error which is within the expected accuracy of the electronic structure approach employed. In addition, the change in the ionization of Tyr57 obtained experimentally upon exchanging H for D (7\%) is in good agreement with the value predicted from our simulations (4.1\%). The under-prediction of the isotope effect on fractional ionization from our simulations is in line with the slightly low value of the simulated excess isotope effect, consistent with recent observations that the B3LYP-D3 density functional slightly underestimates the degree of proton sharing, and hence isotope effects, in hydrogen bonded systems upon including nuclear quantum effects \cite{wang2014}.

The large degree of proton sharing with the deprotonated Tyr57 residue upon including nuclear quantum effects can be elucidated by considering the potential energy required, $\Delta E_{\nu=0}$, to move a proton in the KSI tyrosine triad from its energetic minimum to a perfectly shared position between the two tyrosine groups ($\nu=0$). $\Delta E_{\nu=0}$ depends strongly on the positions of the residues comprising the triad, and in particular on the separation between the proton donor and acceptor oxygen atoms.  Figure 4a shows $\Delta E_{\nu=0}$ computed as a function of the distance between O16 and O57, $R_{OO}$, for the tyrosine triad in the absence of the protein environment. Removing the protein environment allows us to examine how changes in the triad distances from their positions in the enzyme affects the proton delocalization behavior without introducing steric overlaps with other active site residues (see SI Methods Id). 

Figure 4a shows that for the range of oxygen distances observed in the tyrosine triad ($R_{OO}$ = 2.50 -- 2.65 \AA) $\Delta E_{\nu=0}$ for H16 is 3-6 kcal/mol. This is 6-12 times the thermal energy ($k_B$T) available at 300 K, leading to a very low thermal probability of the proton shared state (lower than $e^{-6}=2\times10^{-3}$). However, upon including nuclear quantum effects the system possesses ZPE, which in this system is $\sim$4 kcal/mol. The ZPE closely matches $\Delta E_{\nu=0}$ and thus floods the potential energy wells along the proton sharing coordinate (Fig. 4b) allowing facile proton sharing (Fig. 3), i.e. inducing a transition to a LBHB type regime where the protons are quantum mechanically delocalized between the hydrogen bonded heavy atoms \cite{Cleland1994,Frey1994,Zhao1996}. This leads to qualitatively different behavior of the protons in the active site of KSI$^{D40N}$: {\it{from classical hydrogen bonding to quantum delocalization.}} The proton delocalization between the residues allows for ionization to be shared among three tyrosines to stabilize the deprotonation of Tyr57, leading to the large observed $pK_a$ shift relative to the value in the classical limit (Fig. S2). This change in proton behavior gives rise to the large excess isotope effect since an O--D stretch possesses a ZPE of $\sim$3 kcal/mol, which is no longer sufficient to fully flood the potential energy well in the proton sharing coordinate $\nu$. As the O--O separation is decreased below the values observed in KSI's tyrosine triad, $\Delta E_{\nu=0}$ becomes negligible compared to the thermal energy ($\sim$0.6 kcal/mol at 300 K). Hence, at very short distances ($<$ 2.3 \AA) thermal fluctuations alone permit extensive proton sharing between the residues and the ZPE plays a negligible role in determining the protons’ positions. Thus one would expect a small isotope effect. On the other hand, at bond lengths in excess of 2.7 \AA, $\Delta E_{\nu=0}$ becomes so large ($>$ 8 kcal/mol, Fig. 4a) that the ZPE is not sufficient to flood the barriers, also resulting in a small expected isotope effect \cite{Mckenzie2014}. The large excess isotope effect in KSI$^{D40N}$ thus arises from the close matching of the ZPE and depth of the energetic well ($\Delta E_{\nu=0}$) which is highly sensitive to the O--O distance. Hence although proton delocalization can occur classically at short O--O distances ($<$ 2.3 \AA), nuclear quantum effects allow this to occur for a much wider range of O--O distances (up to $\sim$2.6 \AA), making delocalization feasible without incurring the steep steric costs that would be associated with bringing oxygen atoms any closer. The distances in KSI's active site triad motif thus maximize quantum proton delocalization, which acts to stabilize the deprotonated residue.

\begin{figure}
\centering
\includegraphics[width=\columnwidth]{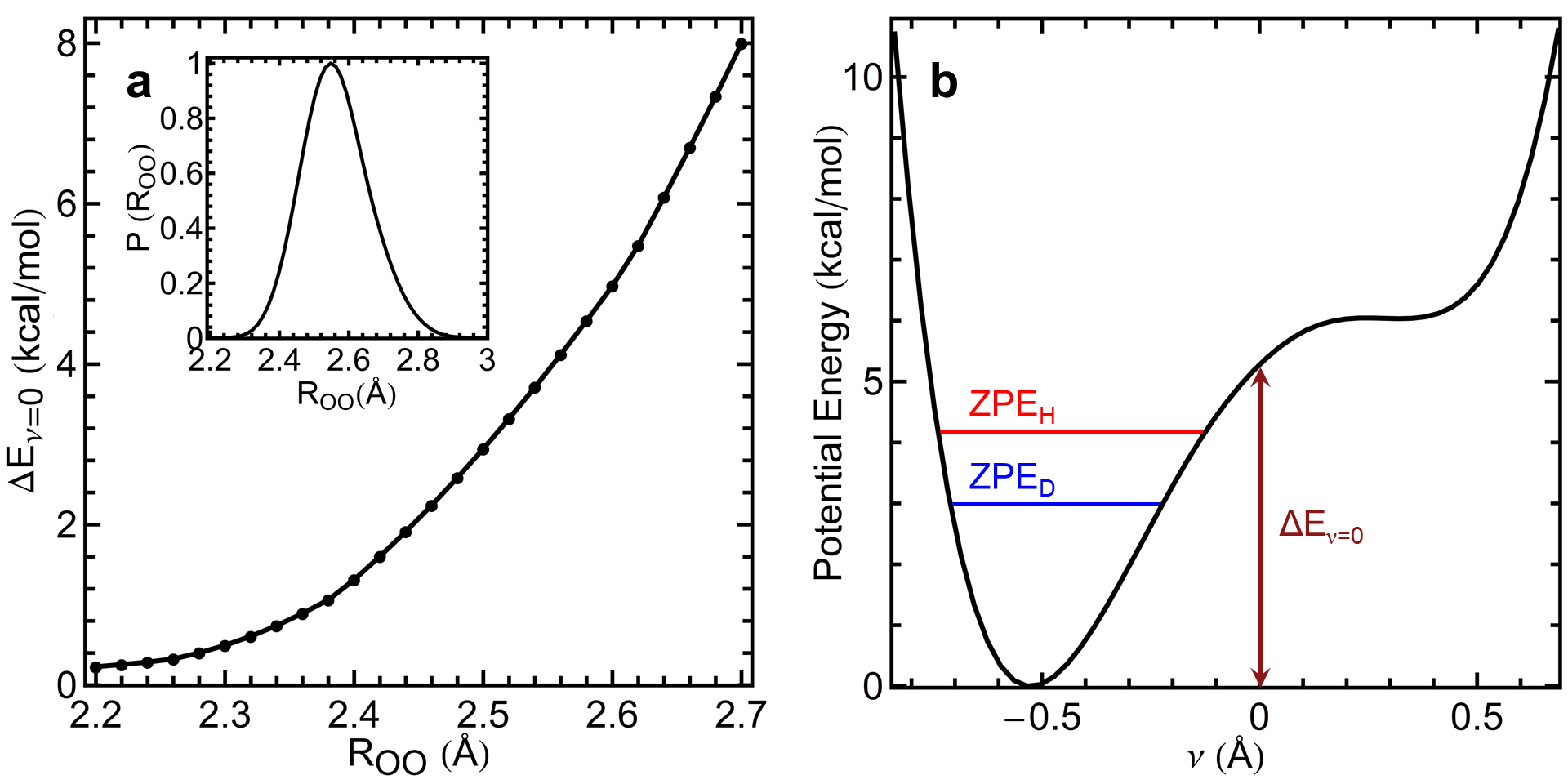}
\caption{Comparison of the energy required to share a proton between residues ($\Delta E_{\nu=0}$) as a function of the hydrogen bond donor-acceptor O--O distance ($R_{OO}$) as compared to the zero-point energy. (a) $\Delta E_{\nu=0}$ as a function of the oxygen-oxygen distance ($R_{OO}$) between O16 and O57 using the tyrosine triad geometry from a crystal structure (details in SI Methods Id). Inset figure shows the probability distribution of $R_{OO}$ obtained from the AI-PIMD simulation of KSI$^{D40N}$ with ionized Tyr57. The probabilities are normalized by their maximum values. (b) Potential energy as a function of the proton transfer coordinate, $\nu$, for $R_{OO}$ = 2.6 \AA ~indicating values for the hydrogen and deuterium (O--H and O--D) zero point energies, $ZPE_H$ and $ZPE_D$. The position of Tyr32 is fixed as the proton H16 is scanned along $R_{OO}$.}
\label{fig:fig4}
\end{figure}

\begin{figure}
\centering
\includegraphics[width=0.6 \columnwidth]{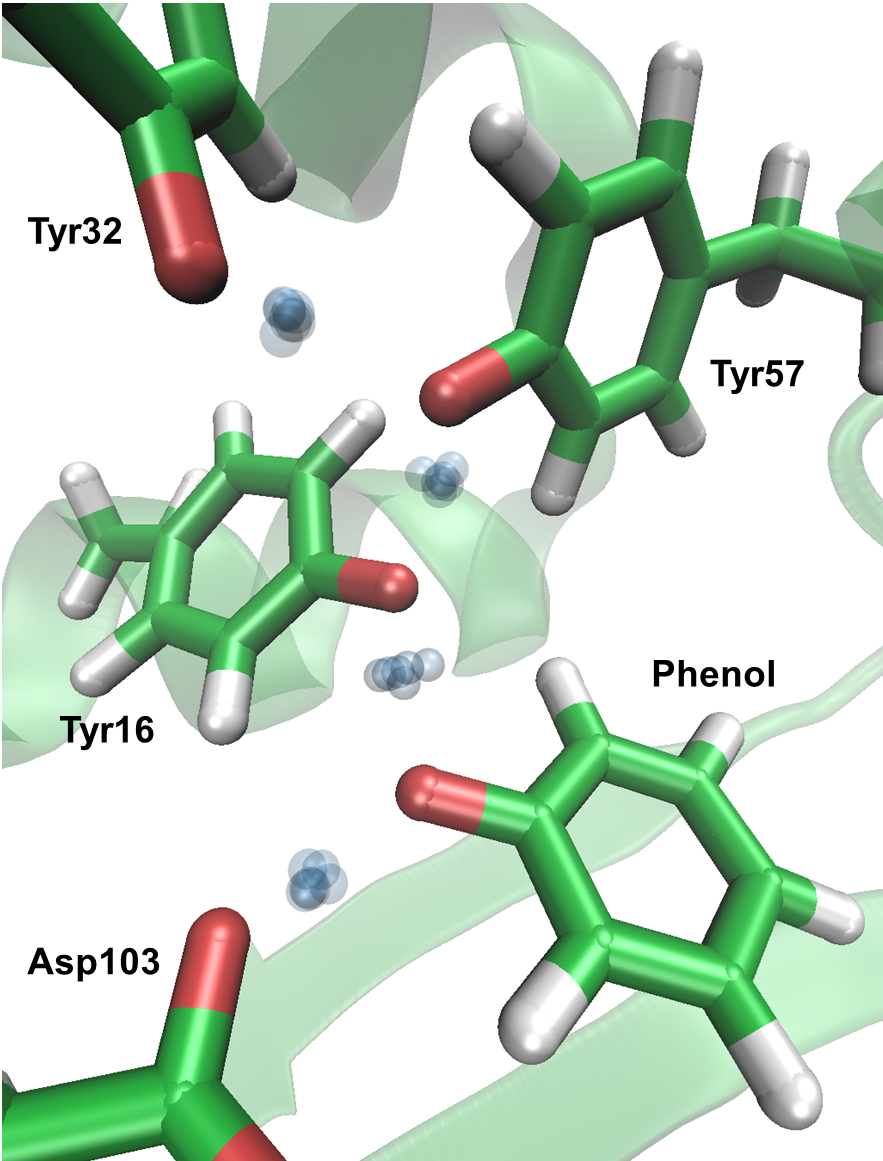}
\caption{Simulation snapshot of the KSI$^{D40N}$ active site with the bound intermediate analog (phenol) that gives rise to an extended delocalized proton network. Green, red and white represent C, O and H atoms. The blue spheres represent uncertainty in the delocalized proton position in the hydrogen bond network. For clarity all other particles are represented by their centroids.}
\label{fig:fig5}
\end{figure}

\subsection{\label{sec:level2}Quantum delocalization stabilizes an intermediate analog by sharing ionization}
Finally, we considered the role of nuclear quantum effects when an intermediate analog participates in the active site hydrogen bond network. Recent experiments have investigated how the binding of intermediate analogs to KSI$^{D40N}$ affects the sharing of ionization along the extended hydrogen bond network that is formed (Fig. 1b) \cite{Sigala2013}. These experiments identified that ionization sharing is maximized when phenol, whose solution $pK_a$ of 10 equals that of the actual intermediate of KSI (Fig. 1a), is bound \cite{Sigala2013,Fried2011,Childs2010}. We thus performed AI-PIMD simulations of the KSI$^{D40N}\cdot$phenol complex. The protons were observed to be delocalized across the network (Fig. 5) with partial ionizations of Tyr57, Tyr16 and phenol calculated to be 18.5\%, 56.7\% and 22.5\% ($\pm$ 0.7\%), compared with estimates from previous experiments using $^{13}$C NMR of 40\%, 40\%, 20\% \cite{Sigala2013}. Hence simulation and experiment are in good agreement that the ionization is shared almost equally among the three residues, i.e. that there is almost no difference ($<$ 2$k_B$T) in the free energy upon shifting the ionization among any of the three groups. Therefore, the ability of protons to delocalize within the KSI$^{D40N}$ tyrosine triad, initially found in the enzyme in the absence of the intermediate analog and manifested as a strongly perturbed $pK_a$, extends upon incorporation of the intermediate analog, which shifts the center of the ionization along the network from Tyr57 to Tyr16. In both cases, proton delocalization acts to share the ionization of a negatively charged group, which suggests that KSI could utilize quantum delocalization in its active site hydrogen bond network to distribute the ionization arising in its intermediate complex (Fig. 1a) so as to provide energetic stabilization.

\section{Conclusions}
In conclusion, KSI$^{D40N}$ exhibits a large equilibrium isotope effect in the acidity of its active site tyrosine residues arising from a highly specialized triad motif, consisting of several short O--O distances, whose positions enhance quantum delocalization of protons within the active site hydrogen bond network. This delocalization manifests in a very large isotope effect and substantial acidity shift. Our simulations, which include electronic quantum effects and exactly treat the quantum nature of the nuclei, show qualitatively and quantitatively different proton behavior compared to conventional simulations in which the nuclei are treated classically, and provide good agreement with experiment. The ability to perform such simulations thus offers the opportunity to investigate in unprecedented detail the plethora of systems in which short-strong hydrogen bonds occur, where incorporating both nuclear and electronic quantum effects is crucial to understand their functions.

\section{\label{sec:level1}Materials and methods}
\subsection{Expression and purification of KSI} Wild-type and D40N KSI from Pseudomonas putida were over-expressed in BL-21 A1 cells (Invitrogen), isolated by affinity chromatography using a custom-designed deoxycholate-bound column resin, and purified by gel-filtration chromatography (GE Healthcare) as described previously \cite{kraut2006}. For $^{13}$C NMR experiments, $^{13}$C$_\zeta$-tyrosine was incorporated into KSI according to methods described previously \cite{Sigala2013}.

\subsection{UV-Vis titration experiments} A series of buffers was prepared with a pL between 4 and 10 by weighing portions of a weak acid and its sodium-conjugate base salt, and adding the appropriate form of distilled deionized water (Millipore H$_2$O, Spectra stable isotopes sterile-filtered D$_2$O ($>$99\% $^2$H)).  Buffers were prepared at 40 mM. Tyr57 is solvent accessible, so the tyrosine residues in the active site network are expected to be fully deuterated in D$_2$O solution.

The following buffer systems were used for the following pL ranges: acetic acid/sodium acetate, 4--5.25; sodium monobasic phosphate/dibasic phosphate, 5.5--8.25; sodium bicarbonate/sodium carbonate, 8.5--10.  Buffers were stored at room temperature with caps firmly sealed.

After preparation of buffers, pL was recorded using an Orion2 Star glass electrode (Thermo), immediately following calibration with standard buffers at pH 4, 7, and 10.  In H$_2$O, the pH of the buffer was taken as the reading on the electrode.  In D$_2$O, the pD of the buffer was calculated by adding 0.41 to the operational pH* from the electrode reading \cite{covington1968}.  A series of samples for titration was prepared by combining 60 $\mu$L protein (100 $\mu$M stock in buffer-free L$_2$O), buffer (150 $\mu$L of 40 mM stock), and extra L$_2$O.  The final samples were 600 $\mu$L, 10 $\mu$M protein, 10 mM buffer.

UV-Vis measurements were carried out on the samples on a Lambda 25 spectrophotometer (Perkin Elmer), acquiring data from 400 to 200 nm with a 1.0 nm data interval and 960 nm/min scan rate and 1.00 nm slit width.  For each measurement a background was taken to the pure buffer of a given pL, before acquiring on the protein-containing sample.  Spectra were recorded in duplicate to control for random detector error.  

The spectra were baselined by setting the absorption at 320 nm to zero, and the change in absorption at 300 nm was followed at varying pL’s, employing a previous established method \cite{schwans2013} to determine the fractional ionization of a tyrosine-tyrosinate pair.  The error in A$_{300}$ from comparing duplicate spectra following baselining was generally between 0--2\%.  For each pL, the average A$_{300}$ was calculated and converted to an extinction coefficient ($\epsilon_{300}$). The titration experiment was repeated on two independently prepared buffer stocks to control for error in buffer preparation.

\subsection{Simulations} AI-PIMD and AIMD simulations were performed using a QM/MM approach of KSI$^{D40N}$ with Tyr57 protonated, KSI$^{D40N}$ with Tyr57 ionized, KSI$^{D40N}$ with the intermediate analog bound, and tyrosine in aqueous solution. The simulations were carried out in the NVT ensemble at 300 K with a time step of 0.5 fs. The path integral generalized Langevin equation (PIGLET) approach was utilized, which allowed results within the statistical error bars to be obtained using only 6 path integral beads to represent each particle \cite{ceri-mano12prl}. The electronic structure in the QM region was evaluated using the B3LYP functional \cite{beck93jcp} with dispersion corrections \cite{grim+10jcp}. The 6-31G* basis set was used as we found it to produce proton transfer potential energy profiles with a mean absolute error of less than 0.4 kcal/mol for this system (see Fig. S6). Energies and forces in the QM region and the electrostatic interactions between the QM and MM regions were obtained using an MPI interface to the GPU-accelerated TeraChem package \cite{ufimtsev2009,isborn2012}. Atoms in the MM region were described using the AMBER03 force field \cite{duan2003}, and the TIP3P water model \cite{jorg+83jcp}. The simulations were performed using periodic boundary conditions with Ewald summation to treat long-range electrostatic interactions. The energies and forces within the MM region, and the Lennard-Jones interactions between the QM and MM regions were calculated by MPI calls to the LAMMPS molecular dynamics package \cite{plim95jcp}. The QM region of KSI$^{D40N}$ contained the p-methylene phenol side chains of residues Tyr16, Tyr32 and Tyr57 (Fig. S5a). For KSI$^{D40N}$ with the intermediate analog, residue Asp103 and the bound intermediate analog were also included in the QM region (Fig. S5b). The QM region of tyrosine in solution contained the side chain of the tyrosine residue and the 41 water molecules within 6.5 \AA ~of the side chain O--H group. All bonds across the QM/MM interface were capped with hydrogen link atoms in the QM region \cite{Eurenius1996}. These capping atoms were constrained to be along the bisected bonds and do not interact with the MM region.

The initial configuration of KSI$^{D40N}$ was obtained from a crystal structure \cite{Ha2000} (PDB ID 1OGX). For KSI$^{D40N}$ with the intermediate analog, a crystal structure \cite{Sigala2013} (PDB ID 3VGN) was used with the ligand changed to phenol.  The crystal structures were solvated in TIP3P water and energy minimized before performing AI-PIMD simulations. The initial configuration for tyrosine in aqueous solution was obtained by solvating the amino acid in TIP3P \cite{jorg+83jcp}water using the AMBER03 force field \cite{duan2003} and equilibrating for 5 ns in the NPT ensemble at a temperature of 300 K and pressure of 1 bar. Each system was then equilibrated for 10 ps followed by production runs of 30 ps.

To calculate the excess isotope effect, $\Delta\Delta pK_a$, (SI Methods Ic) we used the thermodynamic free energy perturbation (TD-FEP) path integral estimator \cite{ceri-mark13jcp}. Combined with an appropriate choice of the integration variable to smooth the free energy derivatives \cite{ceri-mark13jcp}, this allowed us to evaluate the isotope effects in the liquid phase using only a single AI-PIMD trajectory. Simulations performed with D substitution showed no change within the statistical error bars reported.

\begin{acknowledgments}
T.E.M. acknowledges support from a Terman fellowship, an Alfred P. Sloan Research fellowship, a Hellman Faculty Scholar Fund fellowship and Stanford University start-up funds. L.W. acknowledges a postdoctoral fellowship from the Stanford Center for Molecular Analysis and Design. This work used the Extreme Science and Engineering Discovery Environment (XSEDE), which is supported by National Science Foundation grant number ACI-1053575 (project number TG-CHE140013). S.D.F. would like to thank the NSF predoctoral fellowship program and the Stanford Bio-X interdisciplinary graduate fellowship for support.  This work was supported in part by a grant from the NIH (Grant GM27738 to SGB). The authors are extremely grateful to Christine Isborn, Nathan Luehr and Todd Martinez for their assistance in interfacing our program to the TeraChem code.
\end{acknowledgments}

\end{document}